\documentclass[smallextended]{svjour3}
\smartqed  
\usepackage{graphicx}
\usepackage{amsmath}
\usepackage{algorithmic}
\usepackage{array}
\usepackage{float}
\usepackage[export]{adjustbox}
\usepackage{adjustbox}
\usepackage{subfig}
\usepackage[T1]{fontenc}
\usepackage{mwe}
\usepackage{subfig}
\usepackage{float}
\usepackage[noadjust]{cite}
\usepackage[utf8]{inputenc}
\usepackage{listings}
\usepackage{xcolor}
\usepackage{comment}
\usepackage{fixme}
\usepackage{lineno,hyperref,lscape}
\modulolinenumbers[5]
\usepackage{color}
\usepackage{soul}

\fxsetup{
	status=draft,
	layout=inline,
	margin=false,
	theme=color,
	mode=multiuser
}

\begin{document}

\title{Ring-Mesh: A Scalable and High-Performance Approach for Manycore Accelerators}

\titlerunning{Ring-Mesh: A Scalable and High-Performance Approach for Manycore Accelerators}

\author{Somnath~Mazumdar \and Alberto~Scionti}

\institute{S. Mazumdar \at
                Dept. of Information Engineering and Mathematics \\
                University of Siena, Italy \\
              \email{mazumdar@dii.unisi.it}
	    \and
           A. Scionti \at
            LINKS Foundation \\ 
            Turin, Italy \\
            \email{alberto.scionti@linksfoundation.com}
}


\maketitle

\begin{abstract}
There are increasing number of works addressing the design challenges of fast, scalable solutions for the growing number of new type of applications. Recently, many of the solutions aimed at improving processing element capabilities to speed up the execution of machine learning application domain. However, only a few works focused on the interconnection subsystem as a potential source of performance improvement. Wrapping many cores together offer excellent parallelism, but it brings other challenges (e.g., adequate interconnections). Scalable, power-aware interconnects are required to support such a growing number of processing elements, as well as modern applications. In this paper, we propose a scalable and energy efficient Network-on-Chip architecture fusing the advantages of rings as well as the 2D-mesh without using any bridge router to provide high-performance. A dynamic adaptation mechanism allows to better adapt to the application requirements. Simulation results show efficient power consumption (up to $141.3\%$ saving for connecting 1024 cores), 2x (on average) throughput growth with better scalability (up to 1024 processing elements) compared to popular 2D-mesh while tested in multiple statistical traffic pattern scenarios.
\end{abstract}

\section{Introduction}
\label{sec:introduction}
The recent growing interest helps to fuse the machine learning (ML) techniques with traditional HPC approaches~\cite{liu2016application,kurth2017deep} as well as Cloud-based services (known as machine learning as a service --MLaaS\footnote{https://azure.microsoft.com/en-us/services/machine-learning-studio/}$^{,}$\footnote{https://cloud.google.com/products/machine-learning/}). Traditionally, HPC applications are either compute or communication-centric. However, there is no easy way to categorise the traffic generated by the applications in the interconnect subsystem of chip-multiprocessors (CMPs)~\cite{Barrow2009}. Thus, the performance of a network subsystem can be evaluated using synthetic statistical traffic patterns~\cite{Dally2004p}. At run-time, threads communicate at different levels. The flow of data they generate depends on several architectural factors (e.g., the type electrical or optical) and topology of the interconnection, the presence of distributed memory banks, the number of levels in the cache hierarchy). The huge amount of data exchanged among the processing elements (PEs) started to push the limits of traditional interconnections with the growing adoption of large multithreaded applications. For instance, Deep learning~\cite{Lecun2015deep} (DL -- a subset of ML) algorithms are characterised by huge inherent parallelism. Each layer of artificial neurons has to process a large amount of input data following a dataflow style. The way this computation is performed generally reflects into a massive number of parallel threads that can quickly add stress on the interconnection. It reflects the traffic patterns that are quite different compared to other application domains. To achieve higher performance, data exchange inside the interconnect needs to be optimised. Furthermore, the bandwidth usage, latency and energy cost are the primary performance related concern, while taking into account emerging physical limitations (e.g., the way the heat is removed from the chip). In fact, inefficient interconnects may reduce the overall system performance, while consuming a significant portion of the area and power budget of the chip~\cite{Hoskote20075}. Past research works primarily focused on improving the PEs' micro-architecture (i.e., compute element with some local storage), while only a little research has been carried out on the architecture of the interconnections. In fact, most of the hardware accelerators use specialised buses, mesh-based interconnects, crossbars, or their combination~\cite{chen2017eyeriss,Ausavarungnirun2016case,du2015shidiannao,akopyan2015truenorth,Liu2015imr,chen2014diannao,Ravindran1997p}. Although mesh-based interconnections offer a good trade-off between performance, power consumption and available bandwidth, their scalability to a large core count remains limited.

Moving from multicore to manycore~\cite{Bohnenstiehl2016}, the probability of resource contention dramatically increases. Thus, the amount of conflict-free resource sharing inside the chip should be maximised to reduce power cost and also to improve the performance. Although scalable interconnections offer a significant amount of (shared) communication resources to all PEs, the increase in the number of connected nodes up to hundreds (or even thousands) makes the possibility of resource contention still an open issue. Further, it may quickly defeat the advantages of substantial parallelism provided by the higher core count. To unleash the full capability of today's and future CMPs, as well as accelerating DL/ML applications, efficient interconnects must offer high bandwidth, low latency, (possibly) memory coherency support, and also better I/O integration. To this end, Networks-on-Chip (NoCs) built around effective topologies demonstrated to be a possible solution for implementing massively parallel processors (thanks to the advantages offered compared to other alternatives regarding wiring area and power cost~\cite{Lee2007chip,Bolotin2004}). For a small number of PEs, ring topology demonstrated to be very useful, requiring low-radix routers. Interestingly, a ring topology can outperform mesh topology for workloads exhibiting moderate to high memory locality accesses~\cite{Ravindran1997p}. Rings also have been used in commercial systems (e.g., the first generation of Intel Xeon Phi co-processor uses a dual ring topology, while second generation, as well as Intel SkyLake-SP Xeon processor, use a ring-based mesh~\cite{intel_ring1,intel_ring2,horro2019simulating}). The main advantage of using the ring topology is the low latency offered to the data packets to reach their destinations. However, rings also suffer from low bandwidth when connecting a large number of nodes, and soon becomes difficult to scale to hundreds of cores (due to their limited bisection bandwidth). Efforts have been made to connect local and global rings via ``bridge routers'' to improve the scalability together with performance and better energy consumption~\cite{Ausavarungnirun2016case,Hamacher2001h,Vranesic1995n}. On the other side, due to its scalability and the highest level of fault tolerance, the 2D-mesh topology has become very popular to implement the interconnect (e.g., TILE~\cite{Bruce2007}, Polaris chip~\cite{Hoskote20075}, and SkyLake-SP~\cite{intel_ring2}). However, 2D-mesh topology suffers from space and power trade-offs for connecting a vast number of PEs~\cite{Hoskote20075,Vangal20078,Balfour2006design}, which is also verified during our experiments. To overcome these issues, proposals combining 2D-mesh with rings have been presented~\cite{Bourduas2007h,Ausavarungnirun2016case}. Recently, power efficiency also became a major concern for NoC designs connecting several hundreds of cores inside the chip. A conventional 2D-mesh router using an internal crossbar switch can consume the most substantial portion of the power budget~\cite{Hoskote20075}. For instance, the NoC for the MIT Raw processor can consume up to 36\% of total system power~\cite{Wang2003pwr}, while Vangal et al. showed that on the Intel TeraFLOPS chip, the NoC uses up to 28\% of tile power~\cite{Vangal20078}. Other experiments have shown that for large core count (i.e., a 256-core-based CMP) conventional 2D-mesh NoC consume up to 45\% of the total energy~\cite{Harting2012}. 

To tackle these above-mentioned issues and to provide a more scalable, high-performance communication medium for very large tiled CMPs (i.e., CMP equipped with hundreds or even thousands PEs), in this work we propose a hybrid on-chip interconnect targeting kilocore-oriented CMPs (up to 1024 PEs in our experiments). Our target application domain is generic. However, the proposed solution could be optimised further to support emerging applications (such as ML/DL-based applications). To this end, the data packet structure along with the micro-architecture of the routers/switches have been tailored to provide a good trade-off between efficiency, performance and flexibility. In fact, while general purpose applications still rely on the support to the standard floating point arithmetic, DL algorithms tend to adopt more compact data types as well as customised ones (such as IEEE 754 compliant half-precision floating point (FP16)). Our solution leverages a highly efficient hybrid topology with the aim of providing high-performance along with low area cost and better power consumption. Unlike other hybrid solutions, in our work, ring and 2D-mesh NoC are combined without using any bridge-router. This hybrid approach can exploit the principle of computing and data locality exhibited by massive multithreaded applications to confine the traffic mostly inside the rings for better traffic management~\cite{Benson2010,Kandula2009} (not shown as it is out of the scope of this paper). Also, using efficient run-time systems (e.g., the Codelet model~\cite{Suettlerlein2013}) can offer the opportunity to exploit such locality easily. Our proposed architecture provides a customised router architecture which processes the traffic and also bypasses it when necessary for better latency improvement. In general, ring topology provides contention-free traffic without consuming a significant amount of power, thanks to its simple architecture. On the other hand, 2D-mesh topology has a large bisection bandwidth but suffers from the large diameter. Hence, our approach is to use 2D-mesh interconnect for high-speed data transfer between distant PEs, while exploiting rings for data exchange between PEs that are close to each other. The proposed NoC design could be a good candidate for serving communications in modern accelerators as it is agnostic to the specific PE's architecture (we can assume each PE is implementing an in-order execution, and exposing a small local scratchpad memory). Such architecture can be reflected and mapped on modern reconfigurable hardware devices (FPGAs), which may offer ``enough'' resources to implement complex computing systems equipped with hundreds of specialised cores (e.g., coarse grain reconfigurable architectures -- CGRA). Saving hardware resources used by interconnecting logic dramatically contributes to the reduction of the overall power and area cost. The efficiency of the proposed NoC architecture allows using a minimal amount of hardware resources, which contributes to energy saving and also can help to reduce the area cost.  Finally, by extending the interconnect capabilities to adapt to the application requirements dynamically, it is possible for the proposed design to improve performance further, while still reducing power consumption.

In this paper, our contribution can be summarised as follows. \textit{i)} We detail our scalable ring-mesh based NoC design and also evaluate our hybrid topology by implementing it on FPGAs. \textit{ii)} We show how our approach can scale well (from 16 to 1024 PEs) and also outperform the standard 2D-mesh NoC with regards to lower latency, higher throughput and better power efficiency. \textit{iii)} We also discuss its reconfigurability feature and how such a design could be effectively exploited to support specific application domains (e.g., DL/ML applications).

\section{Related Work}
\label{sec:related}
In past years, NoC has received much attention from the research community. Whereas some of the works focused on proposing low latency router (e.g.,~\cite{Kumar2007ex,Hoskote20075}) and power efficient microarchitectures (e.g.,~\cite{Moraes2004h,Wang2003pwr}), other researchers focused on proposing different topologies (e.g., Dragonfly~\cite{Kim2008}, Flattened butterfly~\cite{Kim2007flattened}).~\cite{Ausavarungnirun2016case,Das2009d,Bourduas2007h} tried to improve the performance-power consumption trade-off through the introduction of hierarchical NoC topologies.

\textit{Ring and mesh-based approaches:} HiRD is a hierarchical ring-based NoC design for improving energy efficiency, where buffers within individual rings are not used~\cite{Ausavarungnirun2016case}. It provides buffer support between different levels of the ring hierarchy, and upon the saturation of buffers, flits are deflected in the rings. It needs four levels of hierarchy to connect 256 PEs, while the connection among different levels is based on dedicated bridge routers.~Thus makes it less area and power efficient. CSquare proposes a way of clustering routers so that clusters adopt an internal tree-like organization~\cite{Zheng2015c}. It is a topology with clusters forming a global parallel-oriented structure to provide high scalability. The authors also showed that the proposed topological design improves throughput, while lowers the average latency over mesh-like topologies (under the uniform traffic pattern). In~\cite{Liu2015imr} a multi-ring NoC design has been proposed. In this work, the primary idea is to avoid mesh routers, while an evolutionary algorithm determines every packet's source to destination route. Transportation network inspired NoC (tNoC) is another proposed hierarchical ring topology~\cite{Kim2014tran}. It employs hybrid packet-flit, credit-based flow control mechanism for better scalability, as well as priority-based arbitration for achieving better performance. tNoC allocates channels with a flit granularity, while buffers are allocated with a packet granularity for reducing buffer counts. Koohi et al. proposed 2D-HERT, a hierarchical expansion of a ring topology focusing on optical NoCs~\cite{Koohi2011a}. Kilo-NoC is a topology-aware QoS-oriented architecture, adopting a low-diameter topology~\cite{Grot2011kilo}. It provides a service guarantee for applications with reduced power and area costs. It reduces the extent of hardware support to portions of the die, which in turn reduces router complexity to support large core counts. In~\cite{Bourduas2007h} the authors present a hybrid architecture where a large 2D-mesh is partitioned into several smaller sub-meshes. Next, the sub-meshes are connected using a hierarchical ring interconnect for delivering global traffic. In this work, a bridge module is used for driving traffic to the different levels of the hierarchy. The addressing and routing scheme has also been modified to support the proposed topology.~In~\cite{Ravindran1997p} authors proposed a hybrid NoC subsystem by connecting the hierarchical ring and mesh. Here, multiple rings are connected via special ring interfaces. Unlike our work, the PEs are connected to local rings and global rings connect only local rings (no PEs).

\textit{Other approaches:} In~\cite{Besta2018slim}, Slim NoC (SN) has been proposed, which is primarily based on the Slim Fly topology. The proposed NoC framework generates on-chip layouts with the help of graph and number theory.~\cite{Choi2017chip} proposed a hybrid (i.e., a combination of wired and wireless links) NoC architecture for heterogeneous (i.e., supporting CPUs and GPUs) computing domain. In this paper, authors also claimed that such hybrid topology is energy-efficient for convolutional neural networks (CNNs) training workloads. In~\cite{kwon2017rethinking} the authors proposed an NoC design tailored for the spatial neural network (NN) accelerators which are based on an array of reconfigurable micro-switches. Such micro-switches can be configured to provide fast communication paths from PEs to global memory. Similar to our proposed solution, the adoption of micro-switches with a simple micro-architecture allows consuming very low power. The solution assumes a generic configuration for the PE micro-architecture, while the NoC topology can be optimised for the specific class of NNs. However, the assumption on the overall chip organization (i.e., all PEs reside on one side of the architecture that is opposed to the global buffer), may strongly limit its applicability to other ML algorithms or even to applications belonging to different domains. In~\cite{Das2009d}, a two-tier hierarchical topology consisting of local networks managed through a bus, and a global network controlled by a low-radix mesh router has been proposed. Authors showed that the proposed topology could reduce the latency, power consumption and energy-delay product only for localised communication-based applications.~GigaNoC~\cite{Puttmann2007giganoc} is another hierarchical NoC design which is tailored for specific SoC designs. It uses a packet-switched wormhole routing algorithm for the packet transfer and also supports a parameterizable flit frame. Concentrated mesh (CMesh) is a modified mesh architecture with replicated sub-networks where express channels are used to incorporate the second network without increasing the die area and wire length~\cite{Balfour2006design}. This approach aims at reducing the hop count and load imbalance. Here, the channel length is kept short to reduce energy dissipation, while express channels are used to improve energy efficiency.~In~\cite{Leng2005implementation} authors propose a cluster-based hierarchical NoC architecture where global and local routers are used to manage traffic inside and among clusters of resources. The usage of traditional routers (for all the hierarchy levels) makes it less area and power efficient. 

Here, we propose a hybrid topology based on local rings globally connected through a 2D-mesh. The primary insight behind choosing ring and 2D-mesh for this proposed hybrid topology is that rings can outperform other interconnects regarding performance, power savings and area cost when the number of PEs is low. Furthermore, ring can keep such features over different traffic patterns, as generated by applications belonging to various domains (e.g., multimedia, ML/DL). On the contrary, with a relatively high number of connected nodes, 2D-mesh topology still provides the best trade-off between performance and scalability. Hence, in the proposed solution, the global mesh is intended to route the ``global traffic'' between distant PEs, while rings offer better performance and energy efficiency in managing ``local'' traffic among cores located close to each other. This feature allows our architecture to be scalable with the increased core count. We modified the architecture of the mesh router together with the ring packet transfer mechanism. We avoid the use of bridge routers to connect the local and global network which helps our approach to not require the introduction of further hierarchical levels (unlike~\cite{Ausavarungnirun2016case}) for connecting larger networks (thus contributing to power and area saving). We compared the performance and the power-area costs of our solution with a flattened 2D-mesh topology, since it represents the most adopted NoC topology~\cite{akopyan2015truenorth,chen2014diannao,du2015shidiannao}.

\section{System Overview}
\label{sec:system}
On-chip packet-switched interconnects (i.e., NoCs) provide the physical substrate used by PEs to communicate to each other and also to the primary memory. Besides, NoCs can be extended to support off-chip communications, thus easing the creation of multi-chip computing systems which further extend the design scalability. The main reason behind developing hybrid topologies is to exploit the fact that most of the communication in a parallel application affects a group of resources (i.e., PEs and routers) that are close each other~\cite{Das2009d} in an optimal resource allocation strategy. To this end, efficient Program eXecution Models (PXM) specifically designed to take advantage of locality of computation can be used to reduce the communication overheads and maximise the efficiency of the system~\cite{Suettlerlein2013}. Hence, the optimisation of the local communication may lead to a substantial improvement regarding data-packet latency, throughput and energy efficiency. 

Our design is based on the observation of how the traffic moves in motorways and takes exits. Once the traffic exits the main motorway, it is injected into small roads where simple traffic management decisions need to be taken. Conversely, more complex decisions and management policies are needed at the level of the global interconnect (main motorways). Similarly, our design provides two levels of communications: global traffic is managed by the customised 2D-mesh routers, while local traffic is injected (or ejected) into small rings. Traffic travelling in these two levels of the hierarchy is decoupled and processed differently to improve the system throughput and reduce the communication latency. In general, mesh routers are more complicated than ring switching stations (i.e., switch modules that are responsible for driving the traffic in the ring or ejecting it towards the mesh). The proposed architecture achieves excellent levels of performance and efficiency by exploiting the fact that the majority of the traffic remains restricted to the rings. Such restriction can be adequately provided by a dedicated run-time or additional hardware support (i.e., fine-grain threads are grouped to form a task that can be forced to execute on a group of PEs connected by a ring)~\cite{7911285}. Such restriction of the traffic to the local communication allows the NoC to exploit more efficient and scalable communication mechanisms, (e.g., rings, compared to traditional flattened 2D-mesh). 

\begin{figure}
\centering
\includegraphics[scale=0.27]{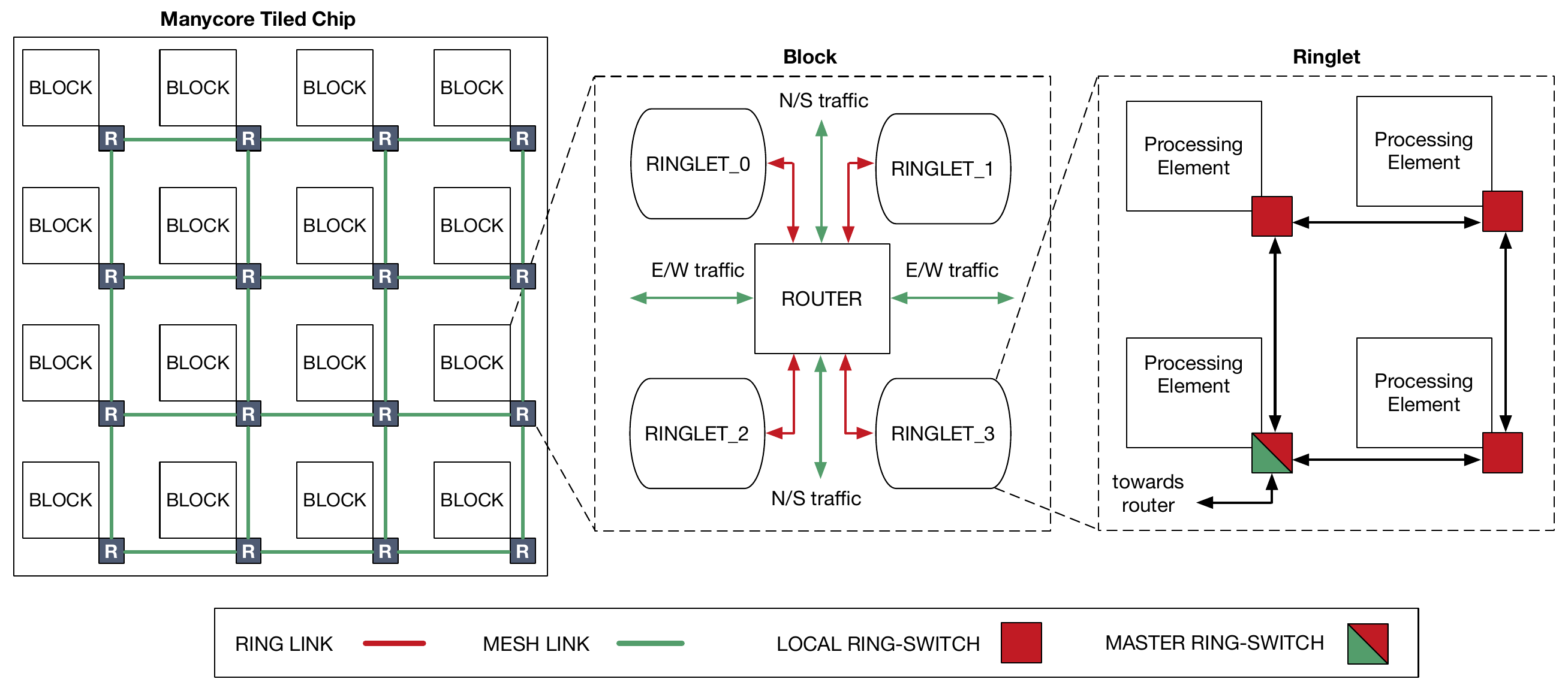} 
\caption{An instantiation of the proposed scalable NoC: 256 PEs organized into $4\times4$ blocks, each connecting 4 ringlets.}
\label{fig:chip-level-architecture}
\end{figure}

The proposed architecture is very f{scalable and can support a large number of PEs (in this paper we set the network size to 1024 PEs). It supports} in-order execution and has a local scratchpad memory block. Apart from that it also uses the deadlock-free X-Y dimension order routing (XY-DoR) algorithm. Figure~\ref{fig:chip-level-architecture} depicts the block diagram of our reference CMP. A group of four PEs are locally connected through a small ring called \emph{ringlet}. Within a ringlet, one of the PEs is designated as the master core. It is responsible for injecting/ejecting traffic towards the global traffic channels. To this end, a link between the ring switch and a mesh router is enabled along with dedicated buffers.
The advantage of this architecture is the absence of a dedicated bridge component to connect the mesh and the ringlets. The proposed NoC is divided into \textit{blocks} formed by a group of four ringlets. Such ringlets are directly linked to the mesh router, which is responsible for moving traffic outside the block. For instance, to support 256 cores, we need 16 modified mesh router and 64 ringlets. 
Multiple blocks are globally connected through a 2D-mesh topology. Each router is equipped with a high performance $8\times8$ internal crossbar switch to support smooth traffic transfer (in and out). The smart packet processing implemented in the mesh routers allow decoupling of global and local traffic. Every time the destination of a packet is outside the local block, the packet is forwarded to another mesh router, thus bypassing PEs in the ringlets, and minimising the overall latency.

\section{Proposed Network-on-Chip Architecture}
\label{sec:working}
In this section, we describe the main components of the proposed NoC architecture. Specifically, we provide details regarding the internal organisation of the mesh router and the ring switch (master core).

\subsection{Modified 2D-mesh router}
\begin{figure*}
\centering
\includegraphics[scale=0.45]{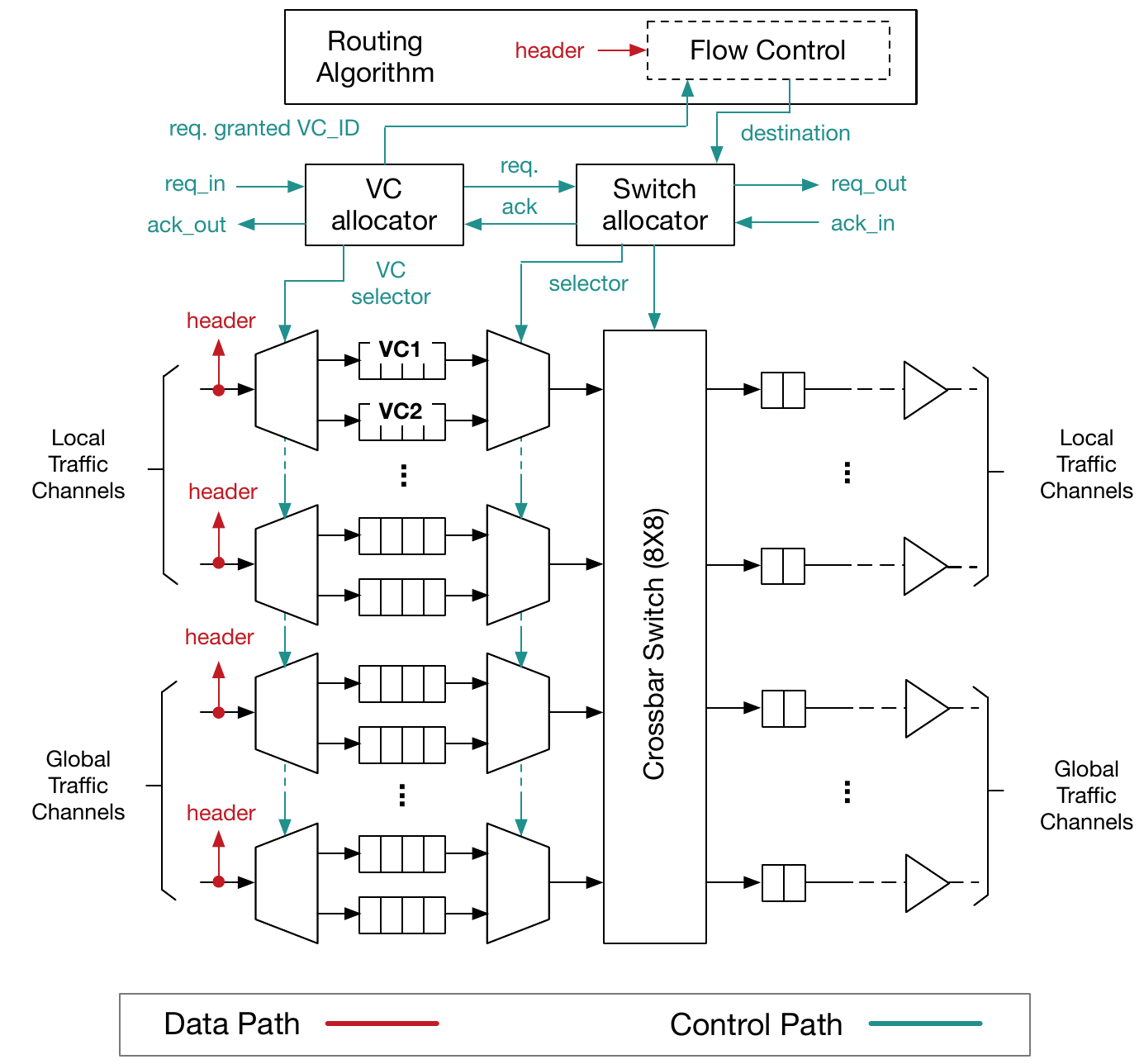} 
\caption{Modified 2D-mesh router micro-architecture: two groups of local/global channels are used to manage traffic within the 2D-mesh and traffic exchange with local ringlets.}
\label{fig:router}
\end{figure*}

Figure~\ref{fig:router} depicts the internal organisation of the mesh router and the Table~\ref{tab:1} provides its main micro-architectural characteristics. The router employs a $8\times8$ crossbar switch to support: \textit{i)} global traffic movement in both dimensions (i.e., North-South and East-West), \textit{ii)} traffic exchange with local ringlets. Four channels are used for driving global traffic within the 2D-mesh network. The other four input channels are used to steer local traffic to/from the ringlets. Each ringlet is associated with a dedicated channel so that the traffic exchange with the master ring switch happens through this dedicated link. In general, routers can have a significant number of virtual channels (VCs) to hold a large amount of incoming traffic while VCs are also used to avoid deadlocks. In Figure~\ref{fig:router}, we highlighted in red the path taken by control information carried by the packet headers and with blue lines the control signals activated by the internal router stages. 
However, large buffer requirements and quality of service (QoS) overheads reduce the ability to support a high number of cores with an efficient area and energy usage~\cite{Grot2011kilo}. In our proposed design, to further saving energy and area, the output channels do not have any VCs because there is less resource contention on the output channels. Furthermore, a large number of VCs consume a huge chunk of energy and also leads to more input buffer counts for traffic management. It is worth to mention that buffers are one of the largest leakage power sources in the router. Their power consumption can represent up to 64\% of the total router's leakage power~\cite{Chen2003leakage} (sometimes it comprises up to 74\% of the total NoC power budget~\cite{Sun2012dsent}), and also a significant amount of dynamic power~\cite{Wang2003pwr}.
In experiments using PARSEC benchmark~\footnote{http://parsec.cs.princeton.edu/}, it has been found that single-flit packets represent the large segment of the network traffic for real applications~\cite{Ma2012whole}. In this work, the proposed mesh-router also supports a single-flit packet (total length of 43-bits) with 32-bits data, and the remaining 11-bits are devoted to carrying header information. The small data packet length represents the bit-stream kind of data transmission, as well as it reflects traffic for the applications that require smaller precision arithmetic and data format (e.g., artificial NNs use low precision data types such as 32/16-bits floating point, 32/16-bits fixed point, 8-bits integer, or even customized ones to represent internal input/output values and the weights). Also, the size of the packet has been chosen by taking into account that increasing the packet size, leads to a quadratic increment of the internal crossbar switch overhead. Thus we maintain the packet size as small as possible~\cite{Lee2013we}, still supporting a wide range of applications. 
\begin{table*}
\centering
\caption{Mesh-router: primary micro-architecture parameters.}
\label{tab:1}
\begin{tabular}{ll}
\hline
\hline
\textbf{Features} & \textbf{Parameters}                                             \\ 
\hline
No. of input and output ports           & 8 each (4 ringlets, 4 mesh)               \\
Width of each port                      & 32-bits (payload) + 11-bits (header) \\ 
No. of Virtual Channel                  & 2 per input port                          \\ 
Packet switching                        & Store-and-Forward (SAF)                   \\ 
Switch allocator arbitration            & Weighted Round-Robin                      \\
Packet Routing                          & X-Y dimension order routing               \\
Router pipeline stages                  & 4 stages                                  \\ 
Latency                                 & 1 cycle (speculation)                     \\
\hline
\hline
\end{tabular}
\end{table*}
The internal router is organised into a standard four-stage pipeline: routing stage, flow-control stage, VC allocation stage, and switch allocation stage. However, with the aim of reducing the latency of the packets during router traversal, proposed design can perform pipeline operations in parallel, thus reducing the overall latency. Thus, an entire packet transfer can be restricted to a single cycle due to
the optimisation of the routing logic for processing the single-flit packet with a reduced overall size. These design choices lead to a router architecture with a latency of one cycle. Our routing mechanism is based on the XY-DoR algorithm since it provides a simple implementation with a deterministic routing latency. Decoupling the traffic between local ringlets and mesh, and exploiting data-computation locality, the probability of congestion in the 2D-mesh is significantly reduced. Thus, the need for an adaptive algorithm (e.g., hot potato routing~\cite{Baran1964dis}, also known as ``deflection routing'') disappear.

\begin{figure*}[ht!]
\centering
\includegraphics[scale=0.45]{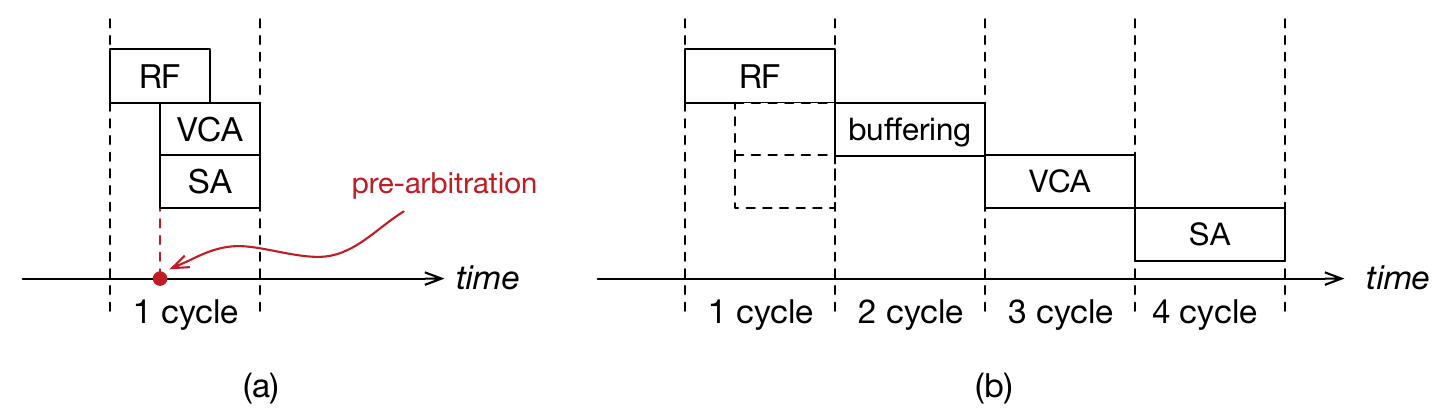} 
\caption{Timing: (a) Best-case: Success of pre-arbitration, (b) Worst-case: Failure of pre-arbitration.}
\label{fig:timing_router}
\end{figure*}
Primarily, we fused the routing logic with the flow-control module. We implemented a speculative allocation technique for both the VC allocation stage (VCA) and the switch allocator stage (SA). In case the pre-arbitration fails, the packet is buffered while VCA and SA arbitration are performed sequentially. In that case, the latency increases up to four cycles. The timing of the proposed mesh router in case of pre-arbitration success is shown in Figure~\ref{fig:timing_router} \emph{(a)}, while the event of failure is represented in Figure~\ref{fig:timing_router} \emph{(b)}. Whenever a packet is entering into the router, the following operations are performed:

\begin{itemize}
    \item \emph{Routing/Flow control module} (RF) extracts the packet header and processes the information to determine the destination router. If the packet destination is within one of the four ringlets belonging to the block, the RF module selects the corresponding output channel, reducing the latency of the VCA and SA module. A control signal is then used to drive the input multiplexer (MUX) at the input port. In this phase, speculative operations are performed to pre-allocate channels.
    \item \emph{VC allocator module} (VCA) is responsible for allocating buffer resources for incoming packets by selecting one of the VCs. An allocation request signal (i.e., \emph{req$_{in}$}) is set, and if the selected VC has space to buffer the incoming packet, an acknowledge signal (i.e., \emph{ack$_{out}$}) is also set. In that case, the selected VC is also signalled both to the RF module and the SA module.
    \item \emph{Switch allocator module} (SA) performs two steps of arbitration. First, multiple VCs in each input port are arbitrated to select one of the available VCs. Next, each one of the selected VCs is routed to the selected output port. 
\end{itemize}

\subsection{Ring switch}
\label{subsec:ring_switch}
\begin{figure*}[ht!]
\centering
\includegraphics[scale=0.415]{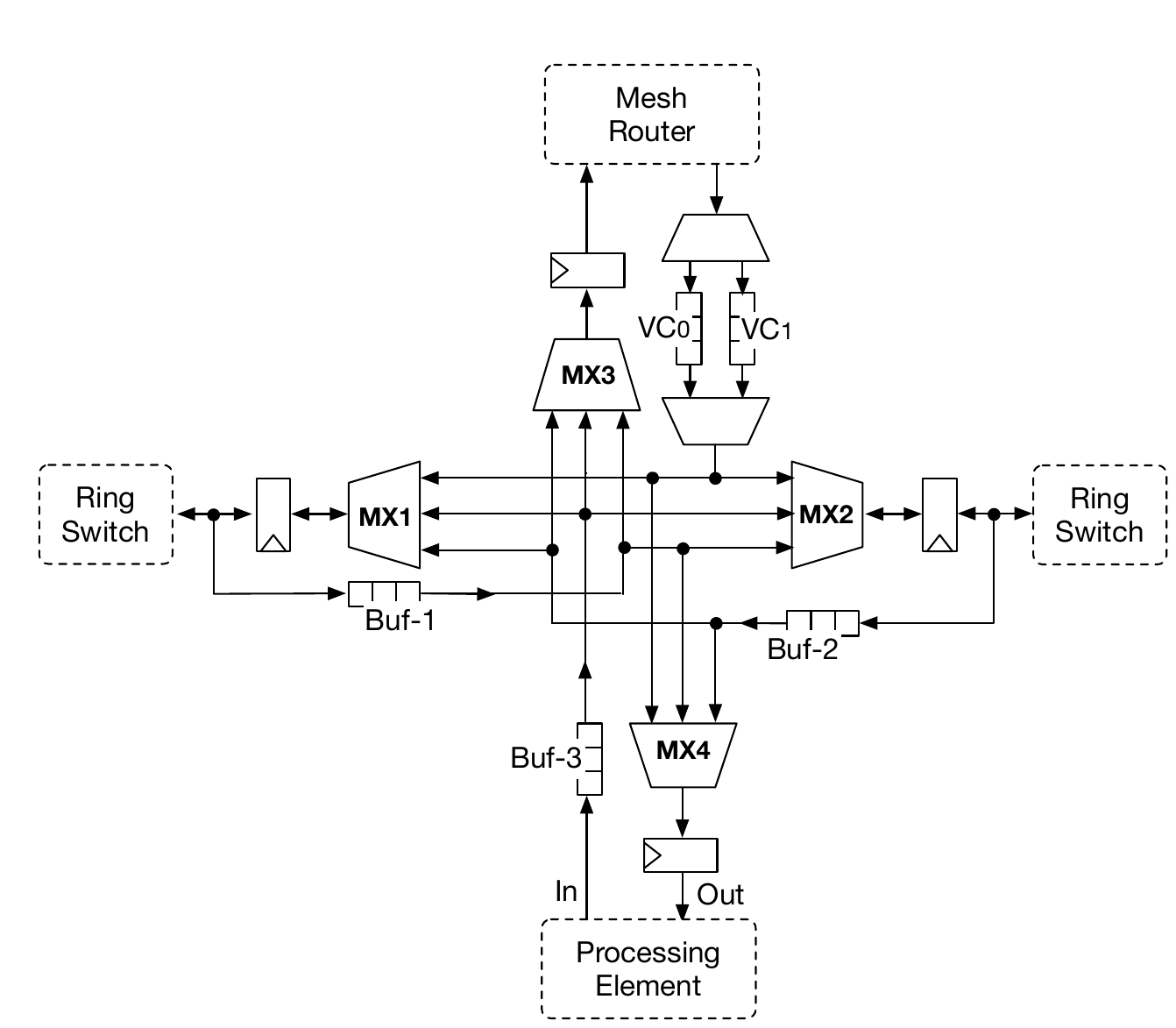} 
\caption{The RS micro-architecture (ringlet master): horizontal dimension is used to create the ring connection, while vertical dimension connects the mesh router and local PE of the ringlet.}
\label{fig:ring}
\end{figure*}

A bi-directional ring is implemented on top of the structure of a \emph{ring switch} (RS) (see Figure~\ref{fig:ring} for microarchitecture of the RS) to achieve better performance while keeping power consumption low. The RS is responsible for driving the traffic within the ring, and also to steer it towards the mesh router. The ringlet uses simple policy to check and forward the flits to the next PE based on the header information. The header contains the indication of the destination ring and the PE responsible for the extraction of the packet from the NoC (see Section~\ref{subsec:data_packet_processing}). In addition to that, a simple modification is done to the existing ring routing policy. In a single ringlet, four PEs are numbering 00, 01, 10 and 11. Packets destined for 00 and 01 will be holding at VC-0 and VC-1 will hold the rests. This simple policy helps to route the packets faster. The RS is composed by two main multiplexers (MX1 and MX2) which manage the traffic within the ring. Compared to conventional RSs, we customised the proposed micro-architecture by incorporating buffers (similar to VCs) to allow the ring to steer the traffic to/from the mesh router. 

To avoid a complex control logic, the RS uses prioritisation of the traffic travelling in the same dimension (i.e., traffic that remains within the ring and moves in the same direction). Prioritization helps to reduce the size of internal buffers too (buffers Buf-1 and Buf-2, see Figure~\ref{fig:ring}). In particular, one of the two directions is selected as with high priority, thus moving first in the RS. Prioritization mechanism is implemented directly in the control logic of input-output multiplexers (MX1 and MX2).
The main drawback of traffic prioritisation is the starvation of the low priority traffic (i.e., buffered traffic is not able to access to the output link), which may potentially wait indefinitely, without winning link arbitration. To avoid such situation, we allow traffic coming from low priority buffers to be injected in the onward link after a fixed amount of elapsed cycles. This mechanism is easily implemented as a slightly modified round-robin selection strategy, where moving from one selected input to another is weighted by the priority of the input. The interface with the local PE is implemented using a dedicated buffer (i.e., Buf-3) which is written by the PE (i.e., the PE injects traffic in the ring) and the RS reads it. The buffer is accessible by the PE within its address space. Similarly, traffic that is ejected by the ring is collected temporarily in a local output buffer, from where the PE can extract the payload. The interface with the mesh router is implemented similarly: traffic injected in the mesh is stored temporarily in a small buffer, from where it is transferred to the input link of the mesh router. Traffic ejected from the mesh router is moved within a VC buffer. When the mesh router tries to access the RS, two VCs are implemented to better support resource contention. From this viewpoint, the RS implements a weighted round-robin selection strategy between the two VCs to keep control logic simple. Such allocation strategy also avoids buffer exhaustion and traffic starvation. Similarly to the PE interface, traffic injection in the router has higher priority, since this reduces pressure on the ring buffers. When packets move within the ring or between a ring and the mesh router, the following steps are performed by RSs:

\begin{itemize}
   \item The multiplexer of each input port (i.e., MX1, MX2, MX3 and MX4 -- see Figure~\ref{fig:ring}) determines the destination based on the packet's header information and also based on the arbitration.
   \item Packets travelling in the ring (i.e., horizontal dimension) have higher priority compared to packets coming from the PE or the mesh router. Thus, packets are moved first from the input port to the output port with a minimal delay. The employed arbitration strategy also ensures that packets already in the main ring traffic flow are quickly routed to prevent the network bandwidth saturation. To enable the transfer, the RS set the request signal of the next switch in the ring (by following the moving direction of the packets), waiting for the acknowledge signal to be set by the peer switch.
  \item The available two VCs' buffers are used to temporarily store the packets when the master RS receives a request from the mesh router to inject packets in the ring. If there is room in the selected VC buffer, the RS enables the corresponding acknowledge signal of the mesh router. Each buffer will take turns to send out the packets via round-robin arbiter to exhibit fairness.
\end{itemize}
   
Worth noting that, RS modules which are not connected to the mesh router have the same structure depicted in Figure~\ref{fig:ring}, except for the mesh router interface to minimise the amount of resources used by routing structures. In that case, the interfaces are removed to save area and power cost. To reduce the pressure on the mesh routers due to the four rings sending packets to each other, the priority rule is applied. In that case, the modified mesh router will process first rings' traffic. This approach avoids the mesh router to become a bottleneck in the communication network. Also, more VCs can be added to the ring but at the expense of the power and hardware cost. Given these features, completing a transaction (i.e., reading a data from a remote PE scratchpad or writing data to) on a fabric block requires not more than 12 cycles.

\subsection{Data Packet Structure and Control Flow}
\label{subsec:data_packet_processing}
\begin{figure}[ht!]
\centering
\includegraphics[scale=0.45]{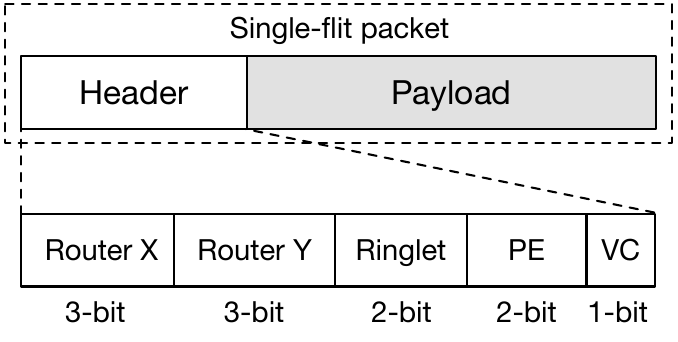} 
\caption{Structure of the single-flit packet.}
\label{fig:pckt}
\end{figure}
The design has been optimized for single-flit packets, and for higher scalability (up to 1024 PEs). Flits have a length of 43-bits. They are formed by a header 11-bits long and a payload of 32-bits. The header structure allows to route the flits within the global 2D-mesh hierarchically and within the ringlets in a block. We have incorporated the XY-DoR protocol in our design. Here, each flits is moved on the X dimension first, and then along the Y dimension, to reach the destination. To support the XY-DoR, the flit header contains two fields, each 3-bits long. In such way, a regular mesh of up $8 \times 8$ routers can be created. Within a physical block (i.e., a router and four ringlets), the ringlets are numbered progressively from 0 (the top-left ringlet) to 3 (bottom-right). Thus, a 2-bits field is used to select the destination ringlet. Similarly, PEs within a ringlet are numbered progressively from 0 to 3. Another 2-bits field is used to select the final destination. Finally, one bit field is used to control the VC assignment. The source router does the assignment of the VC. Aiming at separating traffic generated by applications running on the chip and control traffic (i.e., configuration packets), the latter is assigned to VC-0 by default (see Section~\ref{sec:appl} -- whenever the starting flit of the configuration packet sequence is detected the VC arbitration will select VC-0 for the subsequent flit). This design choice also contributes to simplifying routing and control logic. Figure~\ref{fig:pckt} shows the structure of the single-flit packet used by our proposed NoC architecture. 

The adoption of single-flit packets allows us also to simplify the control flow mechanism. Routers' logic, as well as ring switches manage control signals used to enable packet transfer. This logic use a back pressure mechanism: every time the router/ring switch has to send a flit, it rises a request signal towards the selected next hop. In case the selected router/ring switch (next hop) has no free slots in the input queue to store the incoming flit, it resets the acknowledge signal. Resetting the acknowledge signal allows to temporary stop transmission of flits (packets) at the source node. The complexity of such logic is reduced, although it ensures correctness and deadlock avoidance.

The following steps are followed to steer the data traffic by processing the packet header:
\begin{itemize}
    \item \emph{Traffic confined in the ring}: the block identifier sub-fields are reset, while ring and PE identifiers are set.
    \item \emph{Traffic injected/ejected to/from the 2D-mesh}: all sub-fields are set while the bypass logic is disabled.
    \item \emph{Traffic confined in the 2D-mesh}: the bypass logic is enabled.
\end{itemize}

\section{Target NoC Usage Scenarios and it's Adaptability}
\label{sec:appl}
In this section, we discuss possible extensions of the underlying working mechanism of the proposed network architecture. It has two aims: \textit{i}) better supporting multiple applications with dynamic resource requirements (power consumption reduction or implicit resiliency can be achieved by deactivating/bypassing some components via proposed control packet), \textit{ii}) better network topology adaptation for specific application domains (such as DL/ML applications). For instance, Hadoop MapReduce jobs can vary the number of required PEs during its execution~or the number of artificial neurons and their connections may change moving from one network layer to another. Also, applications built upon an explicit dataflow programming model (e.g.,~\cite{Suettlerlein2013}) show a similar behaviour, where dataflow graphs representing the execution flow and the thread/task dependencies can grow and shrink during the application lifetime. For better processing the different layers of neurons in ML applications, efficient resource mapping on the resource fabric is needed. Intending to support above mentioned applications, we have provided a set of features which could exploit better a large number of PEs. Explicitly, we allow the active elements of the network (i.e., mesh routers and RSs) to be selectively bypassed, completely switched-off or programmed to permit the traffic to be routed through customized topologies.

\subsection{Morphing Capability}
\label{sec:morph}
Smart selection of specific topology to run applications can optimise cost and improve performance. The operating system (OS) or a dedicated run-time can work in conjunction with the NoC support to exploit such feature to optimise the thread/task communication overheads. Here, we present an extension of the proposed design aimed at supporting dynamic reconfiguration of the topology (to better adapt the application needs). The routers and RSs can process a unique configuration single-flit packet called \textit{morph} packet, which specifies how to configure single links within the mesh routers or the RSs. In HPC environments, hypervisor/OS is responsible for generating such control packets consulting the compiler (similar to~\cite{Murali2004su}).
\begin{figure}[ht!]
\centering
\includegraphics[scale=0.33]{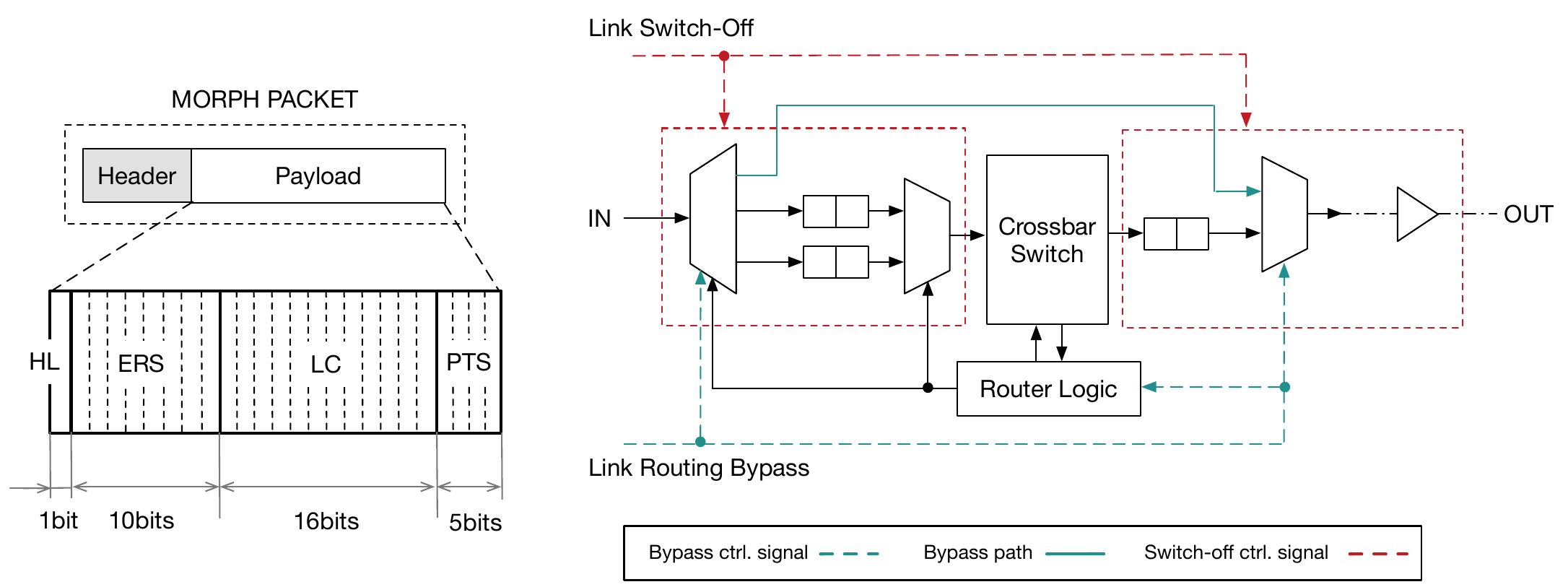} 
\caption{Internal organization of the morph control packet (left), and the corresponding control structures in the mesh router (right). }
\label{fig:morphing}
\end{figure}
Our single-flit packet structure does not permit to encode in the flit header the information for issuing control signals to configure the network. Such information must be carried within the flit payload. We adopted a communication protocol to avoid consuming bits by reserving part of the payload for control information. Here, the control packets are signalled by sending an initial \emph{starting} flit. Since payload with all bits set (i.e., the payload is 0xFFFFFFFF) is rare, we use such value to enable the transmission of the subsequent control flit. Furthermore, a simple combinatorial logic can be used to detect such packets. Conversely, when such data flit must be transmitted to the destination, a pair of flits with the 0xFFFFFFFF payload is transferred. Preliminary simulations confirmed us that the penalty is negligible, when real application traffic is taken into account. To deal with such protocol, the routing logic is slightly augmented with a simple finite state machine that can process configuration packets. The protocol ensures that configuration flits can be correctly delivered without sacrificing bandwidth for the application traffic. Furthermore, configuration traffic can be restricted to use only one of the two VCs available in the routers and RSs (e.g., we can assume to assign VC-0 to manage morph packets). The morph packet is organised in such way the 32-bits of the payload are used to carry configuration information. Notably, the payload is composed of four sub-fields: hierarchy level (HL=1-bit), execution region size (ERS=10-bits), link configuration (LC=16-bits), and PE-type selector (PTS=5-bits). Figure~\ref{fig:morphing} (left) shows the internal organisation of the morph packets.
\begin{itemize}
    \item \emph{Hierarchy level} (HL): Single bit allows distinguishing if the configuration must be applied to an RS (HL = 0) or the mesh router (HL = 1). 
    \item \emph{Execution region size} (ERS): The next 10-bits form this field in the payload. With this length, it is possible for an application demanding for its execution a subset of the total core count, as well as the whole CMP computing resources.
    \item \emph{Link configuration} (LC): This field can use up to 16-bits to specify how to configure single links. Each group of two-bits in this field allows specifying the state of the corresponding link. In the case of the mesh router, links are eight in total (i.e., north, south, east and west for the 2D-mesh and four to connect to the ringlets). Conversely, an RS has four links at most. Thus, only a reduced number of bits in this field are used in the case of RS configuration. A group of two-bits allows to set the link into three main states:
        \begin{itemize}
            \item \emph{Active}: Link is fully active, and the traffic normally flows inside the router/switch for routing decision.
            \item \emph{Bypass}: Link is configured in such way the incoming traffic is directly presented to the corresponding output port, moving in the same direction. For instance, bypassing the east channel of a mesh router allows injecting the traffic directly in west output channel.
            \item \emph{Switch-off}: Link is completely switched off, by disabling the logic governing it. The router/switch logic governing the other links is reconfigured accordingly. 
        \end{itemize}
   
    \item \emph{PE type selector} (PTS): The remaining 5-bits can be used to target particular resources in the chip (e.g., dedicated accelerating cores when a heterogeneous environment is used). To make the system able to distinguish between data payload (0xFFFFFFFF) and a configuration with all the bits set, we force the least significant bit in the PTS field to be set to zero. This design choice still presents enough room to target several types of embedded/IP cores. Indeed, we reserved the configuration 0x00 of the PTS field for identifying further specialized control packets (see Section~\ref{subsec:custom_topologies}).
\end{itemize}

Morph packets can be generated at the OS/hypervisor level in such way they can be sent selectively to a subset of the routers. However, the way control information is organised in the payload, allows the routers and RS not involved in the configuration process to skip their processing, thus avoiding further throughput limitations. Starting from the requests of the application, routers can be organised in such way they create execution regions for each application instance (i.e., an execution region corresponds to the dynamic group of PEs required by the application). For instance, it is possible for a mesh router to dynamically restrict the execution of an application to two ringlets, and use the other two ringlets for the execution of another application (application awareness).
The proposed morphing solution is flexible enough so that it can be exploited to tailor the NoC topology to the application requirements. Similarly to~\cite{Scionti2016}, by selectively bypassing or disabling links, it is possible to allow the NoC to assume a special (virtualised) configuration that provides more performance for the specific application. 

Figure~\ref{fig:morphing} (right) shows the modification to the control signals to allow morphing capabilities. Blue dashed lines represents signals set to bypass crossbar switch and input/output channel buffers. Every time this signal is set, packets entering in the selected input channels are directly driven by the output links (blue line). Conversely, the switch off control signal (it is dominant over the bypass signal) completely disables input and output channel logic. Traffic entering in switched off channels is dropped. Similarly to mesh router, bypass and switch off logic directly operates on the MUXs/DMUXs governing the RSs (see Figure~\ref{fig:ring}, Section~\ref{subsec:ring_switch}). 
 
Morphing capabilities can also be used to improve further power saving (such as power gating technique based on a traffic activity threshold~\cite{Parikh2014p}) and ensure system resiliency. Whenever any running application does not use a portion of the fabric, such resources can be switched off. When an application requires more resources, switched off elements (e.g., a ringlet) can be enabled by resetting switch off control signals. The proposed extension can play an important role to ensure system resiliency to faults. By detecting faulty PEs, or a failure in the router/switch logic, the component can be easily bypassed.

Finally, the morphing capability is mainly aimed at improving the energy-cost by allowing multiple applications to share the spatial as well as computational resources of the chip. The primary idea beyond the use of morphing capabilities is to better tailor NoC resources available on the fabric to adapt to an application context that exploits data and computation locality in a more dynamic way.

\subsubsection{Discussion on Supporting DL Applications}
\label{subsec:custom_topologies}
In~\cite{kwon2017rethinking}, authors show how the generated DNN traffic has unique features that are not captured by well-known benchmark applications. It also suggests to use an NoC structure made of simple switches (similar to our RSs) which can be dynamically reconfigured. Through this capability, the NoC topology could be adapted to support mentioned DL traffic patterns. However, the primary drawback of this solution is the lack of support for traditional traffic patterns. Liu et al.~\cite{liu2018neu} also analysed the performance of a hybrid NoC topology (namely Neu-NoC). To support DL traffic, multiple (local) rings are connected to a global mesh. Such hybrid solution has been demonstrated to outperform the flattened 2D-mesh design by almost 20\%. However, compared to our proposed solution, Neu-NoC architecture suffers from its rigidity. Indeed, the number of PEs that belong to the same ring is fixed.

It is worth to note that this paper reports (in Section~\ref{sec:eval}) the performance of our proposed design based on the traditional traffic patterns only, but we would like to explain the special features of the proposed design for its extended usage purposes (experiments are meant to be for future work). In our proposed NoC design, the traffic flow can be controlled via the morph packets. Such an approach could be ideal to support DL/ML applications. We could map the $0x00$ value carried by morph packets to instruct the destination router(s) to receive further subsequent control flits. The payload of such flits is used to carry information about I/O ports mapping. Such mapping is provided in the form of a table (namely Routing Flow Table (RFT)) of $8\times8$ elements $rft_{i,j} \in \{0,1\}$. If the traffic entering in the input port $i$, can be transferred to the output port $j$ (i.e., $rft_{i,j} = 1$ if the traffic from port $i$ to port $j$ is allowed, $0$ otherwise). To store the table, we could slightly extend the routers' architecture with a small register (such as 64-bit register). Such length is enough to encode binary information contained in RFT. Since flits' payload is 32-bit long, two subsequent flits are enough to carry the whole configuration of the RFT. Every time such custom topologies need to be used, a simple decoding logic of the register content could be used to enable or discharge traffic.

\section{Network-on-Chip Characterization}
\label{sec:noc_characterization}
The proposed hierarchical topology is regular (at each level of the hierarchy) and presents some symmetry, which helps in deriving analytic performance metrics. Here, we consider the two most relevant metrics: the maximum distance, and the bisection bandwidth.
\subsection{Maximum Distance}
\label{subsec:max_distance}
The maximum distance or \emph{diameter} ($\Delta_{max}$) is defined as the maximum shortest path between all pairs of nodes. In a hierarchical topology, the diameter depends on the configuration of each topology level and is proportional to the number of hops to traverse. Diameter determines the worst case distance. Here, the worst case is represented by the communication between nodes placed at the opposite corners of the global mesh. Such nodes reside in two distinct ringlets, for which the diameter is equal to two. It leads us to the following formulation of the topology diameter:
$$
\Delta_{max} = N_R + N_C + 6
$$
Where $N_R$ is the number of links to traverse in the vertical dimension of the global 2D-mesh, $N_C$ is the number of links to traverse in the horizontal dimension of the global 2D-mesh, and $6$ are the links to traverse within the ringlets (three links each). Specifically, since the ringlets are based on bi-directional rings, only two links must be traversed to reach any node inside the rings, plus an additional link that connects the ring to the mesh router.
\subsection{Bisection Bandwidth}
\label{subsec:bisection_bandwidth}
The \emph{bisection bandwidth} ($\beta_{NoC}$) is defined as the bandwidth of the minimum cut that divides the network into two halves. It implies that $\beta_{NoC}$ is equal to the number of links comprised in the cut multiplied by the bandwidth offered by each link. Given our hierarchical topology, regardless of the specific configuration of the two levels, the least connected cut of
the graph representing the network will always be along the global 2D-mesh. Therefore, the bisection bandwidth is given by:
$$
\beta_{NoC} = min(N_R, N_C) \cdot b_l
$$
Where $b_l$ represents the bandwidth of each of the links comprised in the cut. Similarly, it is still possible to determine the bisection bandwidth associated with the other hierarchical levels. Looking at the single block (i.e., a router with four associated ringlets) the bisection bandwidth is given by half of the bandwidth offered by the internal router crossbar switch (i.e., $\beta_{router} = \frac{b_{crossbar}}{2}$); while, the bisection bandwidth of each ringlet ($\beta_{ringlet}$) is equal to that of a bidirectional ring. Although from the above analysis the proposed topology preserves characteristics that are in line with those of well-known topologies (i.e., 2D-mesh and rings). The main advantage resides in the simpler micro-architecture of the rings that provides room for exploring more aggressive clocks without negatively impacting on the overall power consumption. The proposed architecture provides large benefit compared to traditional interconnections adopted in manycore accelerators, along with the exploitation of data and computation locality.

\section{Simulation}
\label{sec:eval}
FPGAs provide a mean to test the effectiveness of the proposed design quickly. Specifically, we synthesised instances of the NoC on a Xilinx Virtex 7 FPGA~\footnote{Model is (XC7VX690 -- speed grade 3)} using Vivado Design Suite 2017.4. All the tests are done setting the clock speed to 400~MHz~\footnote{Although higher clock speed are possible with new generation of FPGA devices.}, and synthetic traffic generators have been developed for generating packets instead of PEs. This section presents the hardware area and power consumption analysis based on synthesis results of the NoC using the FPGA. We simulated the proposed architecture taking into account the average network latency and average throughput. It is worth to note that VHDL model of the network and its synthesis on a real FPGA device allow us to accurately analyse the performance of the proposed design, as well as to measure internal resource and power consumption. Vivado tools have been used to calculate the area and power cost of the design. In the following, we report the primary results by comparing the proposed architecture with a reference design based on a traditional flattened 2D-mesh interconnect. Main micro-architecture parameters for the flattened 2D-mesh design are the same of those used for the proposed modified mesh router, except for the number of input-output ports which is smaller (following a canonical design).

\subsection{Cost metrics}
Cost metrics are represented by the power consumption and the resource utilisation when the proposed design and the flattened 2D-mesh topology are implemented on the FPGA device.

\subsubsection{Resource Utilisation: Area}
We compared two designs on a relative scale and reported the values as the percentage of the total used resources (see Table~\ref{lab:tab3x}). To this end, we have counted lookup tables (LUTs), flip-flops (FFs) and Block RAMs (BRAMs each 36Kb in size). In our proposed design, for the implementation of a single block (i.e., four ringlets connected to a mesh router), the total number of used LUTs, FFs and BRAMs are 2434, 2768 and 48 respectively. Specifically, only four ringlets consume a total of 1076 LUTs, 1800 FFs and 40 BRAMs, while resource consumption of the proposed mesh router is reported in Table~\ref{tab:compare}. 

In Table~\ref{tab:compare}, we also compare the resource utilisation and power consumption between a standard 2D-mesh router with our proposed one. The table reports the absolute number of resources consumed by the two designs, along with the static and dynamic power breakdown. As previously stated, the standard 2D-mesh router share microarchitectural parameters with our proposed mesh router (see Table~\ref{tab:1}), except the number of input-output ports (the number of ports are five in total). Unlike a standard router, the modified one can support sixteen cores via four ringlets with around $2\times$ increment in the resource utilisation compared to a traditional mesh router and with less than $0.4$W increment of power consumption. Regarding the power consumption, we split the values between static power (due to leakage currents and which depends on the manufacturing process) and dynamic power. Results show that the proposed design consume $1.0$ mW and $28.0$ mW more respectively regarding static and dynamic power. The result is strictly correlated to a large number of memory blocks (BRAMs) used by the proposed design.
\begin{table*}
\centering
\caption{Area (resource utilisation) and power consumption comparison between a single 2D-mesh router instance (for single core) and the proposed mesh router (for supporting 16 cores).}
\label{tab:compare}
\vspace*{0.5em}
\begin{tabular}{c|c|c|c|c|c|}
\cline{2-6}
\multicolumn{1}{l|}{}                          & \multicolumn{3}{c|}{\textbf{Area}}                      & \multicolumn{2}{c|}{\textbf{\begin{tabular}[c]{@{}c@{}}Power [W]\end{tabular}}} \\ \hline
\multicolumn{1}{|c|}{\textit{\textbf{Router}}} & \textit{\textbf{LUTs}} & \textit{\textbf{FFs}} & \textit{\textbf{BRAMs}} & \textit{\textbf{Static}}                         & \textit{\textbf{Dynamic}}                         \\ \hline
\multicolumn{1}{|c|}{2D-Mesh}                  & 699                    & 572                   & 5                       & 0.323                                            & 0.047                                             \\ \hline
\multicolumn{1}{|c|}{Proposed}                 & 1358                   & 968                   & 8                       & 0.324                                            & 0.075                                             \\ \hline
\end{tabular}
\end{table*}

We compare resource utilisation of our single block (connecting 16 PEs in total) with the publicly available FPGA friendly NoC generator CONNECT~\cite{Papamichael2012co}. In this experiment, our design saves 74.65\% of LUTs and 39.51\% of FFs compared to CONNECT (we used the code that is available on its official website without any further modification) to connect sixteen cores via an on-chip network. CONNECT also has used 1728 Distributed RAM elements, each 64-bits in size. It is important to highlight that such smaller SRAM elements have a high cost regarding wires when compared to BRAMs.
\begin{table*}
\begin{center}
\caption{Relative resource utilisation: values are expressed as the percentage of FPGA resources consumed. (Note: for the largest network configuration, the resources required exceeds the available resources of a single FPGA device)}
\label{lab:tab3x}
\vspace*{0.5em}
\begin{adjustbox}{max width=13.25cm}
\begin{tabular}[b]{|c||c|c|c|c|c|c|c|}
\hline
& \multicolumn{7}{|c|}{\textbf{System Configuration (No. PEs)}} \\
\hline
 & \textbf{16} & \textbf{32} & \textbf{64} & \textbf{128} & \textbf{256} & \textbf{512}& \textbf{1024} \\ 
\hline
\hline
 & \multicolumn{7}{|c|}{Proposed router design} \\
\hline
\textbf{LUTs}           & 0.31    &0.63    &1.25    &2.51    &5.02    &10.03    &20.06              \\ \hline
\textbf{FFs}            & 0.11    &0.22    &0.45    &0.89    &1.79    &3.58    &7.15                \\ \hline
\textbf{BRAMs}           & 0.54    &1.09    &2.18    &4.35    &8.71    &17.41    &34.83            \\ \hline
\hline
& \multicolumn{7}{|c|}{Ring switch design} \\
\hline
\textbf{LUTs}           & 0.25    &0.50    &0.99    &1.99    &3.97    &7.95    &15.90            \\ \hline
\textbf{FFs}            & 0.21    &0.42    &0.83    &1.66    &3.32    &6.65    &13.30            \\ \hline
\textbf{BRAMs}           & 2.72    &5.44    &10.88    &21.77    &43.54    &87.07    &174.15            \\ \hline
\hline
& \multicolumn{7}{|c|}{Conventional 2D-Mesh router design} \\
\hline
\textbf{LUTs}                 & 2.58    &2.11    &4.23    &20.65    &41.31    &82.61    &165.23        \\ \hline
\textbf{FFs}                  & 1.06    &2.11    &4.23    &8.45    &16.90    &33.80    &67.60        \\ \hline
\textbf{BRAMs}                 & 5.44    &10.88    &21.77    &43.54    &87.07    &174.15    &348.30       \\ \hline
\end{tabular}
\end{adjustbox}
\end{center}
\end{table*}
Scaling the system up to 1024 PEs, we need 64 modified mesh router and 256 ringlets. Such architectural blocks consume up to total 155776 LUTs, 177152 FFs and 3072 BRAM blocks (total of four FPGA devices are used, interconnected each other through dedicated links). From the Table~\ref{lab:tab3x}, we can see that our model is very resource efficient compared to the standard flattened 2D-mesh design. For connecting sixteen cores (one block), our design can save (on an average) 2\% LUTs, 0.7\% FFs and 2.2\% of BRAM compared to the sixteen 2D-mesh routers. Although, it might seem small saving when the design is scaled to 1024 cores the total average resource saving (over all the four FPGAs) increases up to 129.3\% for LUTs, 47.2\% for FFs and 139.3\% for BRAMs.

\subsubsection{Power consumption}
\begin{figure}[ht!]
\centering
\includegraphics[scale=1.0]{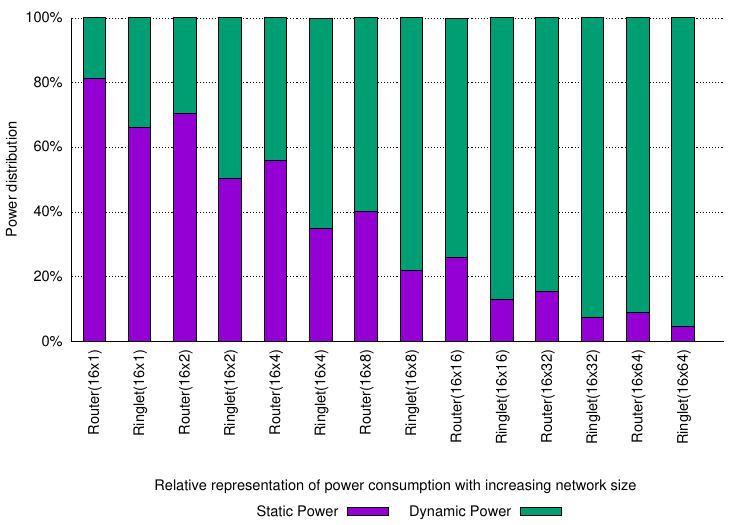} 
\caption{Static and dynamic power breakdown.}
\label{fig:power_dist}
\end{figure}

In Figure~\ref{fig:power_dist}, we show the static and dynamic power breakdown for each network configuration. Reported values are normalised to the total power consumption exhibited by each configuration, and expressed as a percentage. Initially, the static part dominates the power consumption, but as the size of the network grows, it started to diminish. Interestingly, we also notice that the static part slightly grows with the increase in the network size, while the dynamic power component quickly starts to dominate. We can also see from Figure~\ref{fig:power_dist} that the amount of static power is more related to the implementation of the mesh routers (compared to ringlets which confirms the findings mentioned in~\cite{Chen2003leakage}). However, it is worth to note that ringlets in all cases consume more FFs and BRAMs, while router consumes more LUTs (specifically, in the range of 0.06\% to 4.2\% -- see Table~\ref{lab:tab3x}).

\begin{figure}[ht!]
\centering
\includegraphics[scale=1.0]{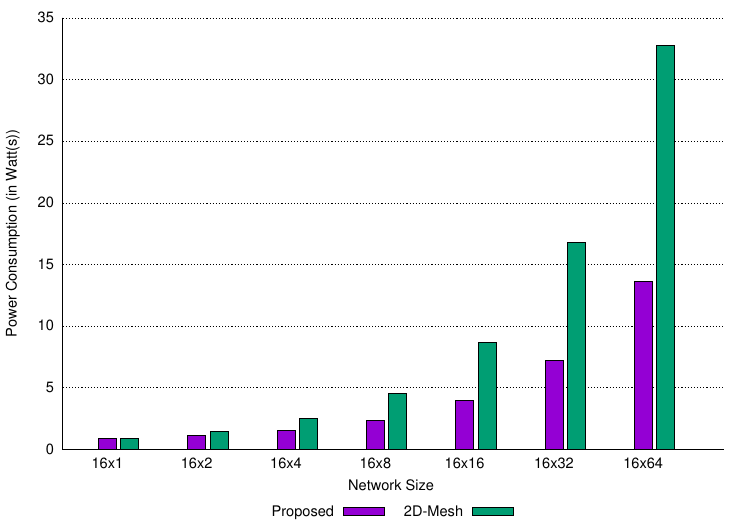} 
\caption{Total power consumption with increasing network size.}
\label{fig:power_tot}
\end{figure}

Next, we plot the power consumption comparison between the proposed model and the standard 2D-mesh in Figure~\ref{fig:power_tot}. It is worth to note that the power consumption of the proposed NoC architecture is the summation of the power consumption of both ringlets and mesh routers. For one topology block (16 cores), the power consumption is $0.399$~W and $0.492$~W, respectively for the mesh router and the ringlet. However, as the size of the network grows, the total power consumption of ringlets starts to dominate. For instance, for 16 topology blocks (i.e., 256 cores), the power consumption of routers is $1.276$~W while the 64 ringlets consume $2.703$~W, which is more than $2\times$ of total routers power consumption. Following this trend, for 1024 core configuration, all the ringlets consume around $2.5\times$ of the total power consumption referred to all the routers. Apart from that, for network size of 16 cores, both the proposed design and the flattened 2D-mesh consume almost the same amount of power. However, as the core count of the network started to grow, the 2D-mesh starts to consume comparatively more power. For the $16\times8$ network configuration, the proposed model consume $2.4$~W while the conventional design consumes $4.5$~W. The situation becomes worse when it touches $32.8$~W for connecting 1024 cores, which represents $141.3\%$ relatively more power compared to the proposed design. Finally, we found that the total power consumption for 16x1 size proposed NoC is 0.89 Watt (0.37 Watt is consumed by the 2D-mesh router). Thus the network for connecting 4x4 cores consume (approx) 58.43\% of the total power. Furthermore, for 16x2 and 16x4 network size, power consumptions are 48\% and 40\% of the total NoC power respectively. Such data confirm that our design is scalable and also power efficient (as the network size grows).

\subsection{Performance metrics}
We evaluated our proposed routers and ringlets under statistical traffic patterns regarding throughput and average data packet latency (i.e., the time for a packet to move from source to destination, including the time for a packet to cross the channel). We consider three well-known statistical traffic patterns (such as uniform random, bit-reversal and transpose~\cite{Dally2004p}) to represent the way real-world applications generate traffic. We developed VHDL-based cycle-accurate models for generating synthetic traffic patterns. In modern computing applications, communication requirements are dynamic and unknown before the execution. Generally, bit-reversal and transpose do not support smooth traffic operations. For the experiment, we generated synthetically a large number of packets that need to be independently routed to a dynamically determined destination. We have used four packet injection rates ($I_r$ -- i.e., it provides the fraction of nodes that simultaneously inject a packet in the network every cycle): specifically $I_r = \{0.25, 0.50, 0.75, 1.00\}$. For $I_r < 1.00$, nodes generating traffic are selected randomly, while with $I_r = 1.00$ (worst case) all the nodes inject packets at the same time. The stress-test was to ensure the working capability of the proposed NoC design under both bandwidth and worst/average latency scenarios.

\subsubsection{Network latency}
\begin{figure}[ht!]
\centering
\includegraphics[scale=0.95]{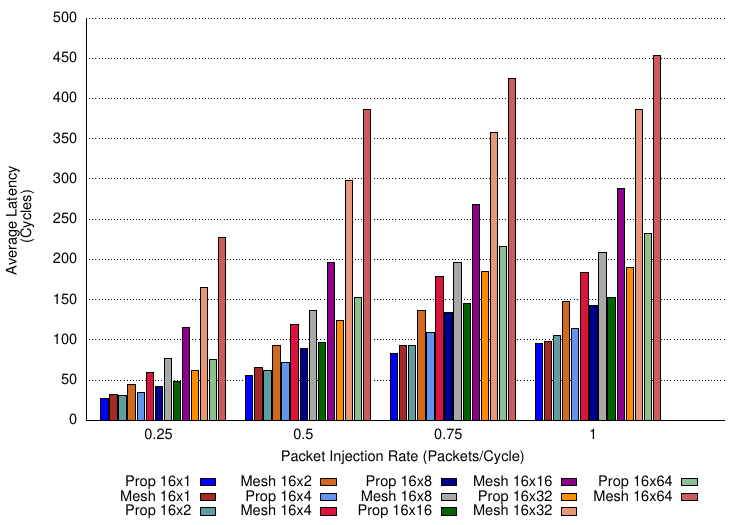} 
\caption{Average packet latency in uniform random traffic pattern.}
\label{fig:trf_uni}
\end{figure}

\begin{figure}[ht!]
\centering
\includegraphics[scale=0.95]{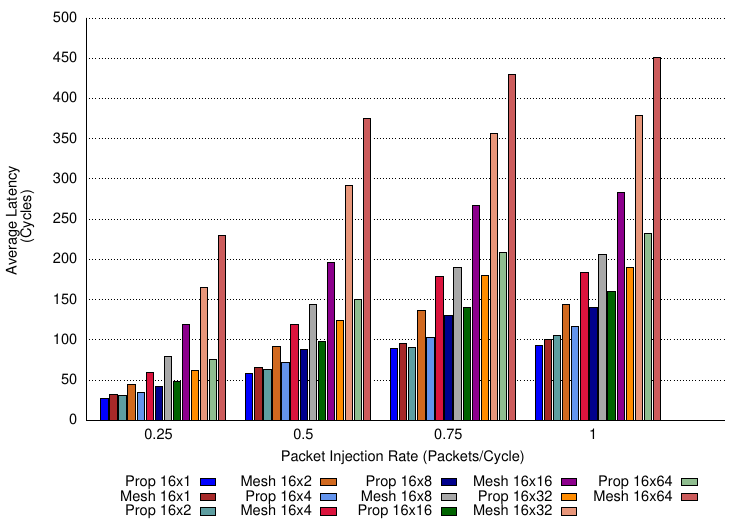} 
\caption{Average packet latency in bit-reversal traffic pattern.}
\label{fig:trf_bit}
\end{figure}

\begin{figure}[ht!]
\centering
\includegraphics[scale=0.95]{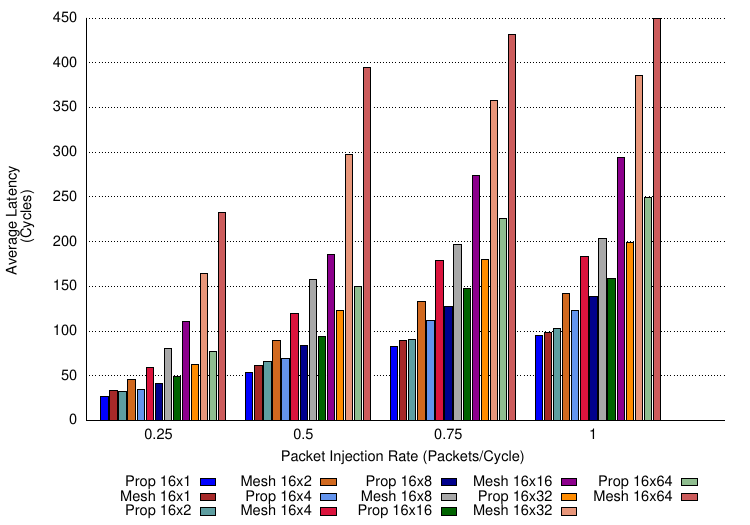} 
\caption{Average packet latency in transpose traffic pattern.}
\label{fig:trf_trans}
\end{figure}

Figures~\ref{fig:trf_uni},~\ref{fig:trf_bit},~\ref{fig:trf_trans} show the average packet latency as a function of the four injection rates, when the three different traffic patterns are used. Bars show that the network latency increases with the increased size of the injection rate, as well as the increase of the network size. However, the proposed system shows very much consistency with increasing network configuration. For low injection rates (i.e., $I_r = \{0.25, 0.50\}$), the latency for each traffic pattern remains very consistent with the others. When increasing the injection rate up to $I_r = 0.75$ and using the bit-reversal traffic pattern, the latency is minimised. The worst case for the packet latency is represented by the transpose traffic pattern with an injection rate of $1.00$. 

When comparing the proposed architecture with conventional flattened 2D-mesh, we found that for all the three traffic patterns, our design outperforms traditional NoC designs, by keeping the latency lower. Specifically, analysing the behaviour of the 2D-mesh NoC, we found that it is very consistent for latency increments for all the cases, while it also has its largest latency for the transpose traffic pattern with the injection rate of $I_r = 1.00$ (similar to the proposed). We improved (on average) the latency by 10\% for all three traffic patterns for smallest network configurations (i.e., 16 cores). While, we improved by $120\%$ for uniform-random traffic, by $115\%$ for transpose traffic, and finally by $124\%$ for bit-reversal for the largest case (i.e., 1024 cores).

\subsubsection{Network throughput}
\begin{figure}[ht!]
\centering
\includegraphics[scale=0.95]{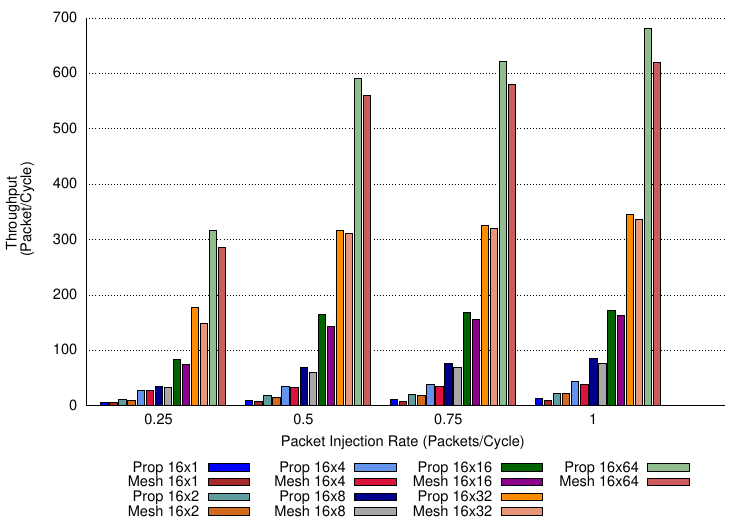} 
\caption{Average network throughput in uniform-random traffic pattern.}
\label{fig:nt_uni}
\end{figure}

\begin{figure}[ht!]
\centering
\includegraphics[scale=0.95]{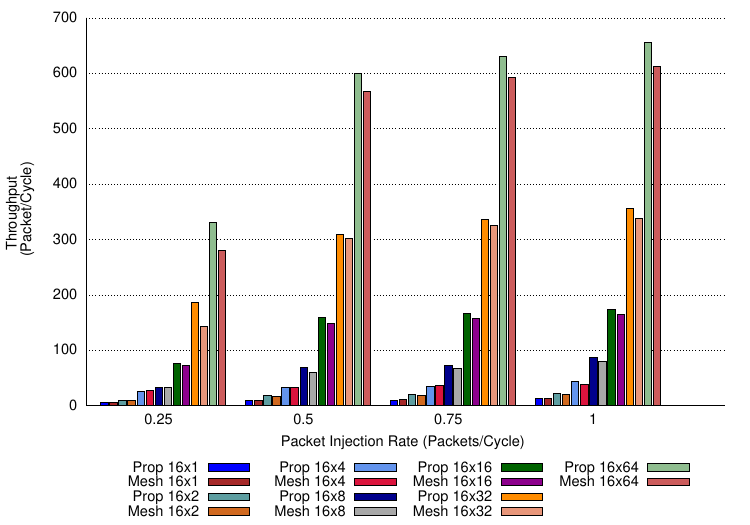} 
\caption{Average network throughput in bit-reversal traffic pattern.}
\label{fig:nt_bit}
\end{figure}

\begin{figure}[ht!]
\centering
\includegraphics[scale=0.95]{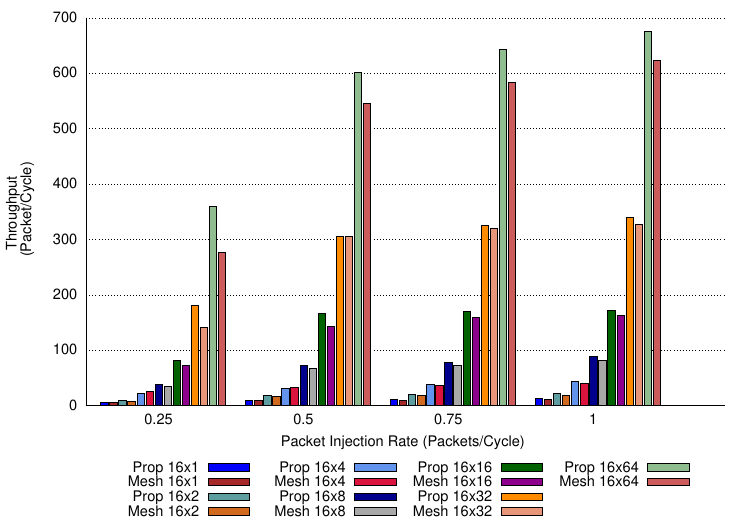} 
\caption{Average network throughput in transpose traffic pattern.}
\label{fig:nt_trans}
\end{figure}

Figures~\ref{fig:nt_uni},~\ref{fig:nt_bit},~\ref{fig:nt_trans} show the achieved network throughput for all three traffic patterns. Similar to latency, the network throughput is also consistent with the packet injection rate. In fact, in the proposed design the average throughput increases as the number of PEs increases. From this viewpoint, by analysing the number of packets delivered per cycle, we observed an increase of a $2\times$ factor with the increase in the number of PEs in the network. For instance, with an injection rate equal to $I_r = 1.00$ and a uniform random traffic pattern, the average throughput increases from 12 packets/cycle for a single block unit (i.e., 16 cores) to 22 packets/cycle for two block units (i.e., 32 cores).

Similarly, for a configuration with 512 cores, the average throughput is 345 packets/cycle, while it increases up to 680 packets/cycle for a 1024 cores configuration. Again, it represents approximately an improvement of a $2\times$ factor with the network size doubling. This trend is also followed by our design when other traffic patterns are considered. It shows that our NoC architecture is capable of offering higher performance and scalability compared to traditional flattened 2D-mesh. Interestingly, we also observed a similar trend in 2D-mesh throughput. Conversely, when the injection rate is low (i.e., $I_r = \{0.25, 0.50\}$), we see that our design performs better for the transpose traffic pattern. For an injection rate equals to $0.75$, the proposed design performed well for all the traffic patterns, while for largest network configuration (i.e., 1024 cores), again the design shows the best throughput for transpose traffic pattern. However, considering the worst injection rate case, it is worth to note that best throughput is achieved with the uniform-random traffic, while traditional 2D-mesh topology did not demonstrate a similar consistency among the patterns. 

These results clearly show that our design can improve the performance of the NoC regarding higher throughput and lower average latency compared to the traditional 2D-mesh topology. The capability of our design to sustain such performance also with high injection rates and random traffic pattern (which represent a critical pattern) can be mainly ascribed to the hierarchical organisation of the network. In fact, most of the traffic is kept inside ringlets or is exchanged by ringlets connected to the same mesh router. Such organisation (ringlet-oriented) is the main contributor to the scalability of the proposed design.

\subsubsection{Network scalability}
\begin{figure}[ht!]
\centering
\includegraphics[scale=1.0]{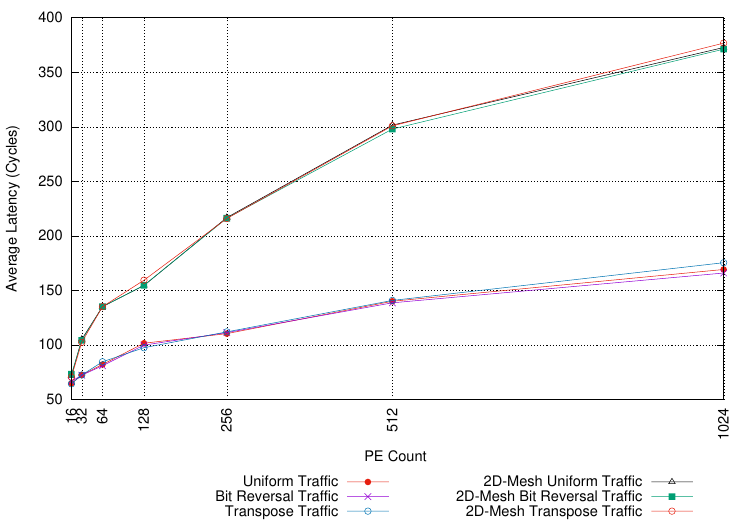} 
\caption{Average packet latency with increasing network size.}
\label{fig:avg_trf}
\end{figure}

To better analyse the scalability of the proposed design, we further performed a set of experiments specifically aimed at evaluating the average packet latency and throughput with the increasing number of cores in the network. For comparison purpose, we averaged the four injection rates (i.e., $I_r = 0.625$) for all the three traffic patterns. The results are shown in Figure~\ref{fig:avg_trf}.

From the plot, it is evident that the proposed NoC architecture shows a significant reduction of the average latency up to 128 cores, compared to the 2D-mesh. Specifically, moving from a configuration with 16 PEs to the that counting 128 PEs, the latency increases from $65$ to $100$ cycles. Conversely, for the same two configurations, the 2D-mesh topology shows an increment of the latency from $72$ to $156$ cycles. This behaviour has been observed irrespective of the traffic pattern. 
For the next two network configurations (i.e., 256 and 512 cores respectively) the latency increment exhibited by our design is better (e.g., the increment from 128 to 256 PEs shows a lower slope of the curve), while in the 2D-mesh the latency increment still follows a linear trend, thus showing worse performance. Finally, considering the most significant configuration (i.e., 1024 cores) the trend is still non-linear for the proposed design and the latency drops to $170$ cycles (this trend is similar for all the three traffic patterns). The trend of average packet latency improvement is also analogous for the other traffic patterns. Conversely, 2D-mesh increases the latency up to $377$ cycles. It is worth to note that, although the two architectures have similar behaviour when the number of the PEs increases, the average latency is always significantly lower with the proposed design. If we combine this observation with the lower power consumption and resource utilisation, we can advocate that the proposed design scales better than conventional ones and it can be a good candidate for supporting next-generation high-performance manycore accelerators. Interestingly, to have similar performance using the flattened 2D-mesh topology, more resources are needed (e.g., a more substantial number of VCs, deeper buffers), leading to a more power hungry and area consuming solution (similar the case reported in~\cite{Hoskote20075,Vangal20078,Balfour2006design}). It is the primary reason that standard 2D-mesh does not scale well with increasing network size.

\begin{figure}[ht!]
\centering
\includegraphics[scale=1.0]{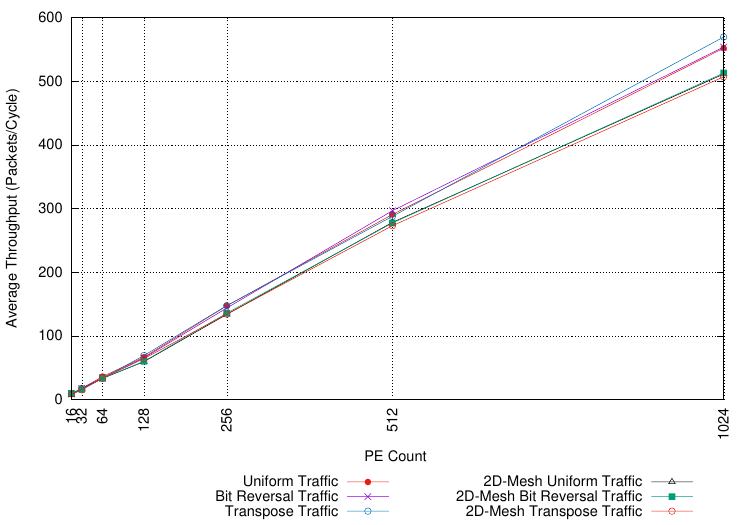} 
\caption{Average network throughput with increasing network size.}
\label{fig:avg_thpt}
\end{figure}

\begin{figure}[ht!]
\centering
\includegraphics[scale=1.0]{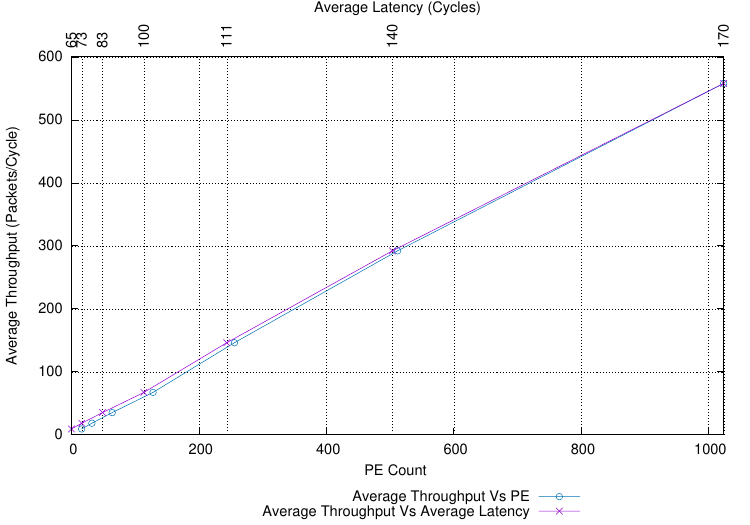} 
\caption{Comparing average network throughput and average packet latency of the proposed design with increasing network size.}
\label{fig:all}
\end{figure}

To further confirm the capability of our design to scale, we also analysed how the average network throughput (i.e., the result obtained averaging the throughput for a given number of PEs to connect) improves with the growing network size (using the same packet injection rate for the three traffic patterns -- $I_r = 0.625$). The results of this experiment are shown in Figure~\ref{fig:avg_thpt}. From the trend, it is evident that the average throughput tends to increase almost linearly by a factor $\approx2\times$ when the number of PEs is doubled. For instance, considering the transpose traffic pattern, for a single block unit (i.e., 16 cores) the average throughput is $9.8$ packets/cycle while it increases to $17.13$ packets/cycle for 32 PEs. Similar behavior is observed when moving from 128 cores ($69.25$ packets/cycle) to 256 cores ($147.7$ packets/cycle), as well as when moving towards the largest configuration, i.e., from 512 cores ($288$  packets/cycle) to 1024 cores ($570$ packets/cycle). Although the 2D-mesh topology shows similar behaviour for lower core counts, it is important to highlight that the average throughput is always lower than our proposed design, and quickly start to decrease when the core count increases (i.e., for more than 128 cores our ring-mesh combination far outperforms the 2D-mesh topology).  
Finally, we compared how the network performs when increasing the number of PEs, by plotting the average throughput versus the number of processing cores and also the average throughput versus the average latency together. The result of this comparative analysis is reported in Figure~\ref{fig:all}. Here, all the reported values of latency and throughput are the average of all three synthetic traffic patterns used in the experiments. From the plot, it appears that the average latency grows by a factor of $1.25\times$ when moving from 256 cores to 512 cores while all the other cases we can see a lower latency growth. The NoC design also registers the average throughput growth of a factor of $\approx2\times$ for almost all cases. The proposed design has shown its robust performance by improving its throughput higher than latency, and even the latency growth is started to reduce with the increase in the network size.
It is worth to note that, good performance exhibited by our design also derive from the adoption of the single-flit packet format. In fact, this format greatly reduces the processing overhead associated with the packet header, since information regarding traffic movement within the ringlets and the global mesh are decoupled. 
%
%

\section{Conclusion and Future Work}
\label{sec:conclusions}
Following the current trend, we will have general-purpose chips with hundreds, or even thousands of PEs. Unfortunately, in the literature most of the proposed solutions targeted the design and optimization of PEs' architecture, not the interconnect subsystem for a potential source of performance gain (particularly in DL/ML-based accelerator domain). To efficiently use the massive built-in parallelism, we need to manage the traffic inside the chip efficiently, both from the power (reducing hot-spots) and performance (lower latency and throughput) perspective. In this paper, we propose a simple yet scalable two-level hybrid hierarchical interconnection where rings and a 2D-mesh topology are fused without using any bridge router. We have implemented the proposed NoC design with an efficient traffic generator on an FPGA device. Using  multiple synthetic traffic patterns, we showed that our design is scalable while keeping high performance regarding throughput and latency. Experimental results also showed that our hierarchical organisation of the interconnect could easily outperform the capabilities of traditional 2D-mesh NoCs. In general, the popular NoC topologies are implemented in hardware accelerators to support the emerging applications (such as DL/ML). This proposed topological design can also be extended using the special configuration packets to exploit chip resources better, depending on the specific application requirements. We also have discussed how the proposed design could be utilised for the applications whose requirements are dynamically changing over their execution lifetime. 

The future work will be to compare the performance of our design against other NoC topologies (such as~\cite{Ausavarungnirun2016case,Liu2015imr,Bourduas2007h,Ravindran1997p}). Furthermore, the quantitative performance analysis of our design will be done while mapping real DL/ML applications. Finally, it will be interesting to explore its performance by exploiting different micro-architectural parameters and larger packet sizes.

\bibliographystyle{spmpsci}
\bibliography{ref}
\end{document}